\documentclass[journal=cmatex,manuscript=article,layout=traditional]{achemso}

\pdfoutput=1 
\usepackage[version=3]{mhchem} 
\usepackage[T1]{fontenc}       
\usepackage{nicefrac}          
\usepackage{color}
\usepackage[normalem]{ulem}
\usepackage{gensymb}

\newcommand{\ai}{\textit{ab-initio}\ }
\newcommand{\clath}[1]{#1$_{8}$Al$_{x}$Si$_{46-x}$}

\newcommand{\spel}[1]{{\usefont{T1}{lmtt}{b}{n} #1}}

\author{Maria Troppenz}
\affiliation{Institut f\"ur Physik and IRIS Adlershof, Humboldt-Universit\"at zu Berlin, Berlin, Germany}
\author{Santiago Rigamonti}
\affiliation{Institut f\"ur Physik and IRIS Adlershof, Humboldt-Universit\"at zu Berlin, Berlin, Germany}
\author{Claudia Draxl}
\affiliation{Institut f\"ur Physik and IRIS Adlershof, Humboldt-Universit\"at zu Berlin, Berlin, Germany}
\alsoaffiliation{Fritz-Haber-Institut der Max-Planck-Gesellschaft, Berlin, Germany}
\email{claudia.draxl@physik.hu-berlin.de}

\title{Predicting ground-state configurations and electronic properties of the thermoelectric clathrates Ba$_{8}$Al$_{x}$Si$_{46-x}$ and Sr$_{8}$Al$_{x}$Si$_{46-x}$}

\begin{document}
\newpage

\begin{tocentry}
\centering
\includegraphics[width=9.0cm]{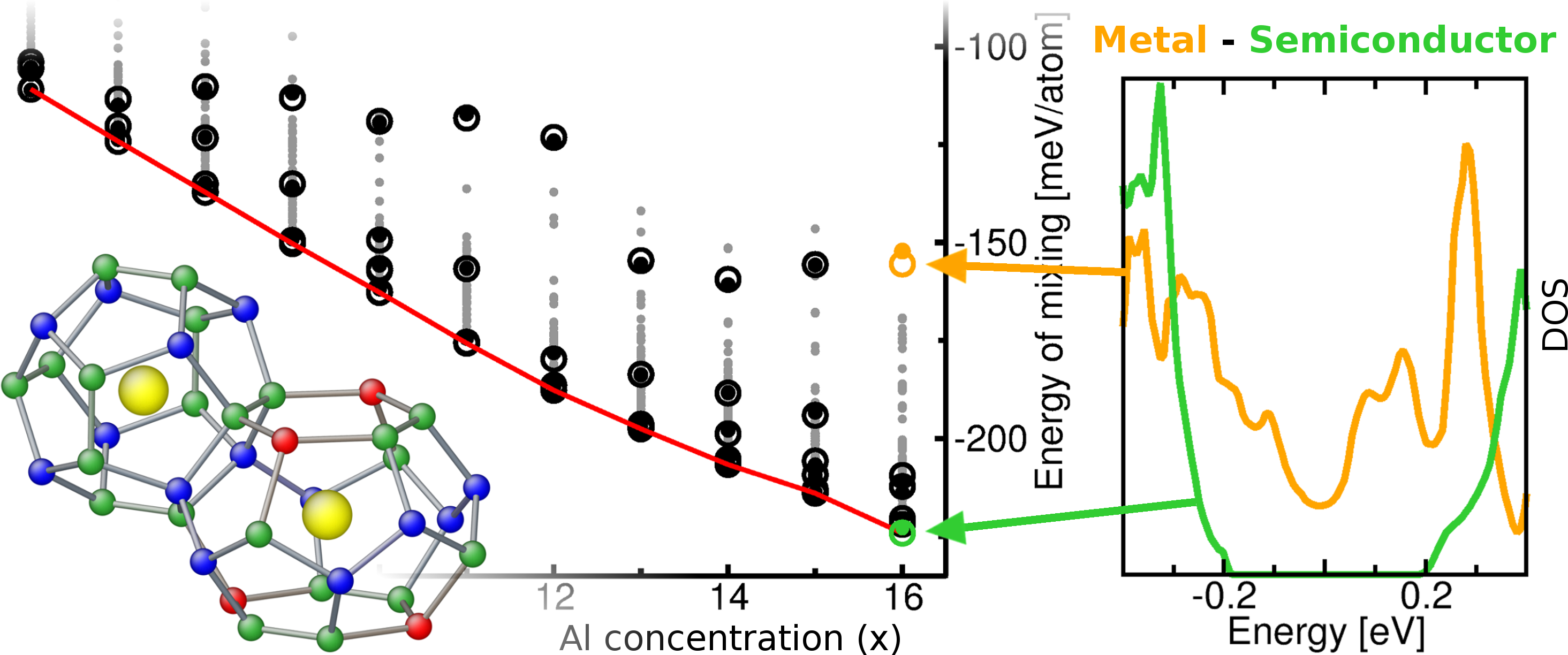}
\end{tocentry}

\begin{abstract}
The structural and electronic properties of the clathrate compounds Ba$_{8}$Al$_{x}$Si$_{46-x}$ and Sr$_{8}$Al$_{x}$Si$_{46-x}$ are studied from first principles, considering an Al content $x$ between 6 and 16. Due to the large number of possible substitutional configurations we make use of a special iterative cluster-expansion approach, to predict ground states and quasi-degenerate structures in a highly efficient way. These are found from a simulated annealing technique where millions of configurations are sampled. For both compounds, we find a  linear increase of the lattice constant with the number of Al substituents, confirming experimental observations for \clath{Ba}. Also the calculated bond distances between high-symmetry sites agree well with experiment for the full compositional range. For $x$ being below 16, all configurations are metallic for both materials. At the charge-balanced composition ($x=16$), the substitutional ordering leads to a metal-semiconductor transition, and the ground states of Ba$_{8}$Al$_{16}$Si$_{30}$ and Sr$_{8}$Al$_{16}$Si$_{30}$ exhibit indirect Kohn-Sham band gaps of 0.36 and 0.30 eV, respectively, while configurations higher in energy are metals. The finding of semiconducting behavior is a promising result in view of exploiting these materials in thermoelectric applications. 
\end{abstract}

\newpage

\section{Introduction}
Waste heat is generated by all kinds of engines, from smartphones and laptops to high-performance computers, from refrigerators to power plants, from cars to airplanes, and many more. Transforming heat into electricity by exploiting the thermoelectric effect provides a possibility to reuse part of this waste heat, however, new materials with large thermoelectric efficiency are a prerequisite for profitable applications. Thus the investigation of and the search for novel thermoelectrics \cite{so2009:lowth,jef2008:TEm,sh2011:TE,ka20141:te,ka2015:te2} concern a hot topic in materials research. 

\begin{figure}[h]
\includegraphics[scale=0.55]{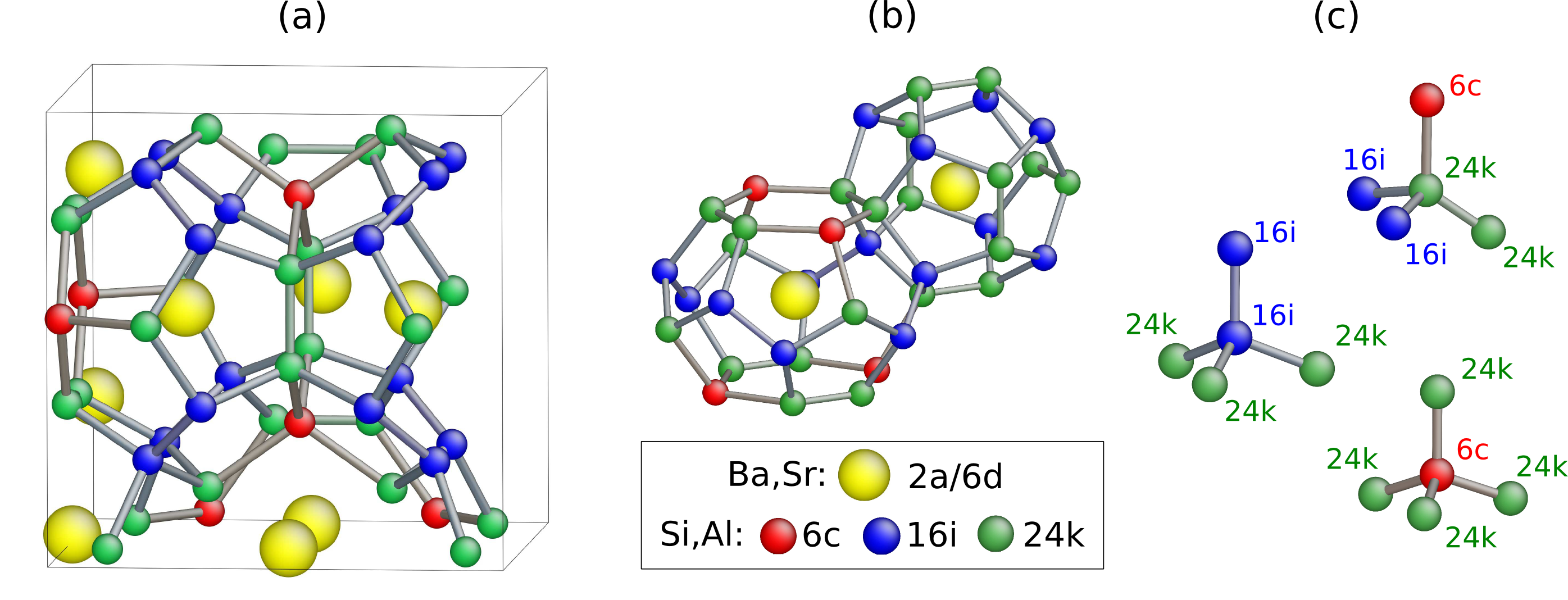}
\caption{Type-I clathrate structure: (a) Unit cell (space group $Pm\overline{3}n$, No. 223) consisting of (b) large tetrakaidecahedral and dodecahedral cages. The Wyckoff sites $24k$ are marked in green, $16i$ in blue, and $6c$ in red, respectively; guest atoms at $2a$ and $6d$ in yellow. (c) Wyckoff sites of the framework and their nearest neighbors.}
\label{fig:struc}
\end{figure}

Intermetallic clathrate compounds \cite{no2001:clath,ko2004:clath,kl2010:clath,chri2010:TE1,to2014:pgec} are promising candidates. Their cage-like structure containing guest atoms is regarded as a realization of the {\it phonon-glass electron-crystal}, promising a large figure of merit. Figure \ref{fig:struc}\,(a) shows the unit cell of type-I clathrates with formula unit G$_{8}$Y$_{x}$X$_{46-x}$ (cubic space group $Pm\overline 3 n$). The most important framework atoms X are group-IV elements, i.e. Si, Ge, or Sn. The host structure Y$_{x}$X$_{46-x}$ is built from 46 atoms forming eight cages, six tetrakaidecahedra and two dodecahedra (Fig. \ref{fig:struc}\,(b)). They occupy the three symmetrically distinct Wyckoff sites $24k$, $16i$, and $6c$, and are interconnected by tetrahedral covalent bonds as depicted in Fig. \ref{fig:struc}(c). This bonding, analogous to that of group-IV semiconductors, can lead to high performance in the electronic properties\cite{cohn1999}. The eight guest atoms G, placed inside the cages at Wyckoff sites $6d$ and $2a$, can induce static and dynamic disorder, thereby lowering the thermal conductivity. They are also called {\it rattlers}, as they are expected to vibrate at low frequencies in a non-directional fashion, efficiently scattering low energy phonons responsible for heat transport.

Without substituents ($x=0$), all valence electrons of the framework atoms X fully contribute to the covalent bonds. Following the Zintl-Klemm concept, the eight guest atoms, G=Ba or Sr considered here, donate their two valence electrons to the host, making the material metallic.  These 16 free charges per formula unit can be compensated by partially substituting the group-IV element X by a group-III element, e.g. Y=Al, Ga, In. Assuming a purely ionic host-guest interaction and complete absorption of the free charges by substituents, full compensation is reached at $x=16$ \footnote{Previous calculations \cite{ro2012:Srclath} of the electron-localizability indicator suggest that, for Al-Si based clathrates, the assumption of an ionic guest-host bonding is indeed reasonable.}. This so-called charge-balanced composition \cite{fae2011:ztl,chri2010:TE1,ki2000:ztl} is expected to be a semiconductor, which is highly desired for thermoelectric applications \cite{eisenmann1986}. 

The variety of chemical elements available for building the clathrate structure offers a wide compositional space for optimizing their electronic properties and reaching high efficiency \cite{co2004:clathbg,shi2010:TEclath}. Clathrates based on Ge and Ga have demonstrated good performance in their thermoelectric properties, \textit{e.g.} in terms of a glass-like thermal conductivity of Sr$_{8}$Ga$_{16}$Ge$_{30}$ \cite{ch2000:SrGe,sa2001:SrGe} and high thermoelectric efficiency of Ba$_{8}$Ga$_{16}$Ge$_{30}$ \cite{to2008:BaGe,sa2006:BaGe}. 
Al-Si-based clathrates are of technological interest in terms of price,
weight, and low environmental impact
\cite{condron2006a,condron2006b,condron2006c,nagatomo2012,anno2013,bo2015:Bax7,ro2012:Baclath,ro2012:Srclath,nen2008:proBaclath,mudryk2002,he2014}. So far, they could be synthesized in the range $x\leq15$ for \clath{Ba} \cite{ro2012:Srclath,bomi2015:pc}, and only for $x=10$ in the case of \clath{Sr}\footnote{The slightly higher electronegativity of Sr as compared to Ba may lead to a lower solubility of Al in Sr-filled clathrates.}\cite{ro2012:Srclath}. Thus, in both cases the charge-balanced composition has not been achieved so far. Still the question is whether the Zintl-Klemm concept holds, i.e. this composition would indeed exhibit semiconducting behavior like the guest-free, non-substituted structure. Yet, it is not clear what the ground-state configurations at this and other concentrations of the substituent for different guest elements are. Only understanding the subtle interplay between covalent bonding and ionic repulsion, tailoring of the material's structure and, hence, its electronic properties will be possible.

In this work, we investigate the clathrate compounds \clath{Ba} and \clath{Sr}. On the search for their ground states as a function of Al substitution in the range $6 \leq x \leq 16$, we explore the configurational space by making use of the cluster expansion (CE) technique \cite{sa1984:cexp}. This technique allows for building numerically efficient models to predict the energy $E$ of a substitutional alloy by exploiting the unique dependence of $E$ on the configuration. Combined with Monte Carlo (MC) simulations, it has been successfully applied to access the stable phases and thermodynamical properties of bulk and surface alloys\cite{mu2003:revc} (for examples, see Refs. \citenum{hue2009:alloyCE} and \citenum{st2005:s}). Due to the large primitive cell of the clathrates, we apply a novel iterative cluster-expansion technique (iCE) \cite{letter}, as implemented in the package \spel{CELL}\cite{cell}. This technique, specially designed for alloys with large parent cells, provides an efficient way to predict the energy of an arbitrary configuration (i.e. a specific arrangement of Al atoms at the lattice sites) with \ai accuracy. Exploring the configuration space, we provide an unbiased prediction of the ground-state configurations as a function of Al content. Based on an in-depth analysis, we discuss their stability and their electronic structure.

\section{Methodology}
\label{sec:methods}
\subsection{Cluster expansion}
\label{sec:methods1}

An arbitrary configuration of the crystal can be represented by a vector ${\bf s}=(\sigma_{s1},\sigma_{s2},...)$, where the occupation variables $\sigma_{si}$ take the value $1$ or $0$, if a site $i$ is occupied by one or another type of atom. For a configuration $s$, the predicted energy $\widehat{E}_s$ can be expanded in terms of $0$-, $1$-, $2$-, ... n-body clusters $\alpha=\{i,j,...\}$ as
 \begin{eqnarray}
  \widehat{E}_s = \sum_{\alpha}\,m_{\alpha} \,J_{\alpha}\, X_{s \alpha},
  \label{eq:Ecluster}
 \end{eqnarray}
where $i$, $j$, ... indicate crystal sites. The coefficients $J_{\alpha}$ are the so-called effective cluster interactions (ECI), and the sum runs over a set of $N_c$ symmetrically distinct clusters. The correlation functions
 %
 %
 %
 \begin{equation}
 X_{s \alpha} = \frac{1}{m_{\alpha}} ~ \sum_{\beta \equiv \alpha} \, f_{s\beta}~
 \label{Ecorrelations}
 \end{equation}
are obtained from an average of the cluster functions $f_{s\beta}$ over all the clusters (denoted $\beta$) which are symmetrically equivalent to the cluster $\alpha$. There are $m_{\alpha}$ such clusters. The cluster functions are defined as the product of the occupation variables $\sigma_{si}$ for the sites $i$ belonging to the cluster, i.e., $f_{s\beta}=\prod_{i \in \beta } \sigma_{si}$. For more details on the CE we refer the reader to Ref. \citenum{sa1984:cexp}. The coefficients $J_{\alpha}$ in Eq.~\ref{eq:Ecluster} are obtained by fitting the predicted energies to a set of $N_t$ structures --the training set-- whose energies are known from \ai calculations based on density-functional theory (DFT)\footnote{Our zero-temperature calculations do not include zero-point vibrational contributions, which could affect to some extent the cluster interactions\cite{garbulsky1994}.}. Formally, this requires the minimization of the objective function
 \begin{eqnarray}
  S^2=\frac{1}{N_t} \sum_{s=1}^{N_t} \, (E_s- \widehat{E}_{s})^2 +  \sum_{\alpha } \, A_{\alpha}  |J_{\alpha}|^2
  \label{eq:CVpenal}
 \end{eqnarray}
with respect to the ECIs, i.e. $\nabla_J \,S^2 = 0$. In Eq.~\ref{eq:CVpenal}, the first term on the right hand side is the mean squared error (MSE) of the predicted energies $\widehat{E}_{s}$. The second term represents an $\ell_2$ regularization which allows for finding an optimal set of ECIs even if the number of clusters $N_c$ is larger than $N_t$. Both the optimal penalization strength $A_{\alpha}$ and the optimal set of clusters are obtained by minimizing the cross-validation score (CV) \cite{st1974:loo,wa2002:atat},
 \begin{eqnarray}
  \textrm{CV}^2=\frac{1}{N_t} \sum_{s=1}^{N_t} \, (E_s - \widehat{E}_{(s)})^2~\textrm{.}
  \label{eq:CVloo}
 \end{eqnarray}
Here, $\widehat{E}_{(s)}$ is the predicted value for $E_s$, which is obtained from a training set excluding the data point $s$. Other optimization procedures as the compressed sensing \cite{ti1996:cs2,ha2013:compsens} consider an $\ell_1$ norm as regularization term, instead of an $\ell_2$ norm as used in Eq.~\ref{eq:CVpenal}. For the clathrate compounds of this study, such an approach did not lead to improvements, neither in the quality of the model nor in the accuracy of the predictions.

In this paper, the CE is performed in an iterative manner, making use of the python package \spel{CELL} \footnote{The code CELL uses the corrdump utility of ATAT for the generation of clusters and the cluster correlation matrix. For details on this utility see Ref.\citenum{walle2009}.}\cite{cell}. It enables cluster expansions for large parent cells, where a full enumeration of possible structures and the corresponding calculations of their predicted values $E$ become an impossible task in terms of computational cost. This is the case for the type-I clathrate compounds. 
Additional details about the iCE approach and its application to complex alloys, as well as the numerical implementation, will be published elsewhere\cite{letter,cell}. 

The search for the ground-state (GS) structures is carried out in an iterative procedure. It starts with a set of random structures, one for each composition $x$ ($6\leq x \leq 16$). The \ai energies of this set are calculated, and a first CE is constructed from them. The large size of the clathrates' parent cell prevents a full enumeration of symmetrically distinct structures for each composition. Instead, the configurational space is explored through a Metropolis sampling generating a Boltzmann distribution at fixed temperature. The energies of the new structures visited in this procedure are predicted by the CE. From this sampling, the lowest-energy structures visited (or the lowest non-degenerate, in case no new ground state is found) are adopted, and their \ai energies are calculated. Based on this extended set of \ai data, a new CE is performed. This procedure is repeated until no new GS structure is found and the CV is small enough to make predictions. 

\subsection{Computational details}
\label{sec:methods2}  
The \ai calculations are performed with the full-potential all-electron DFT package \spel{exciting}, \cite{gu2014:exciting} an implementation of the (linearized) augmented planewave + local-orbital method. In order to evaluate the stability of our results, we use two different exchange-correlation functionals, namely the	local-density approximation (LDA) \cite{pe1992:lda} and the generalized-gradient approximation in terms of the PBEsol functional \cite{pe2008:PBEsol}.

For obtaining a CE with high predictive power, well converged \ai calculations are essential. The crucial convergence parameters, affecting the accuracy of the calculation, are the number of $\mathbf{k}$-points, used for the discretization of the Brillouin zone and the basis set size. The latter is determined by the parameter $R_{\min}\times G_{\max}$, where $G_{\max}$ is the maximum reciprocal lattice vector of the augmented planewaves used in the representation of the wavefunctions, and $R_{\min}$ is the smallest muffin-tin radius. Total energies are calculated with $R_{\min}\times G_{\max}=8$ and a $4 \times 4 \times 4$ {\bf k}-grid, leading to  an accuracy of energy differences between structures of the same composition below $0.07$meV/atom. The uncertainty in the lattice parameter upon volume optimization is smaller than 0.01{\AA}. The atomic positions are relaxed until all the forces are below a threshold of $F_{tol}$= 2.5 mRy/atom. We find that the energy differences scale roughly linearly with the number of Al-Al bonds, where each of them leads to an increase by about 7-10~meV/atom. Energy differences between structures of the same composition and no Al-Al bonds vary by a few meV/atom, with the smallest differences being around 0.2~meV per atom for structures differing only in one Al-Al second nearest-neighbor pair.

\section{Ground-state search}
\label{sec:gss}

\subsection{Results of the iterative cluster expansion}
\label{sec:gss1}

\begin{figure}[h]
  \centering
  \includegraphics[scale=0.57]{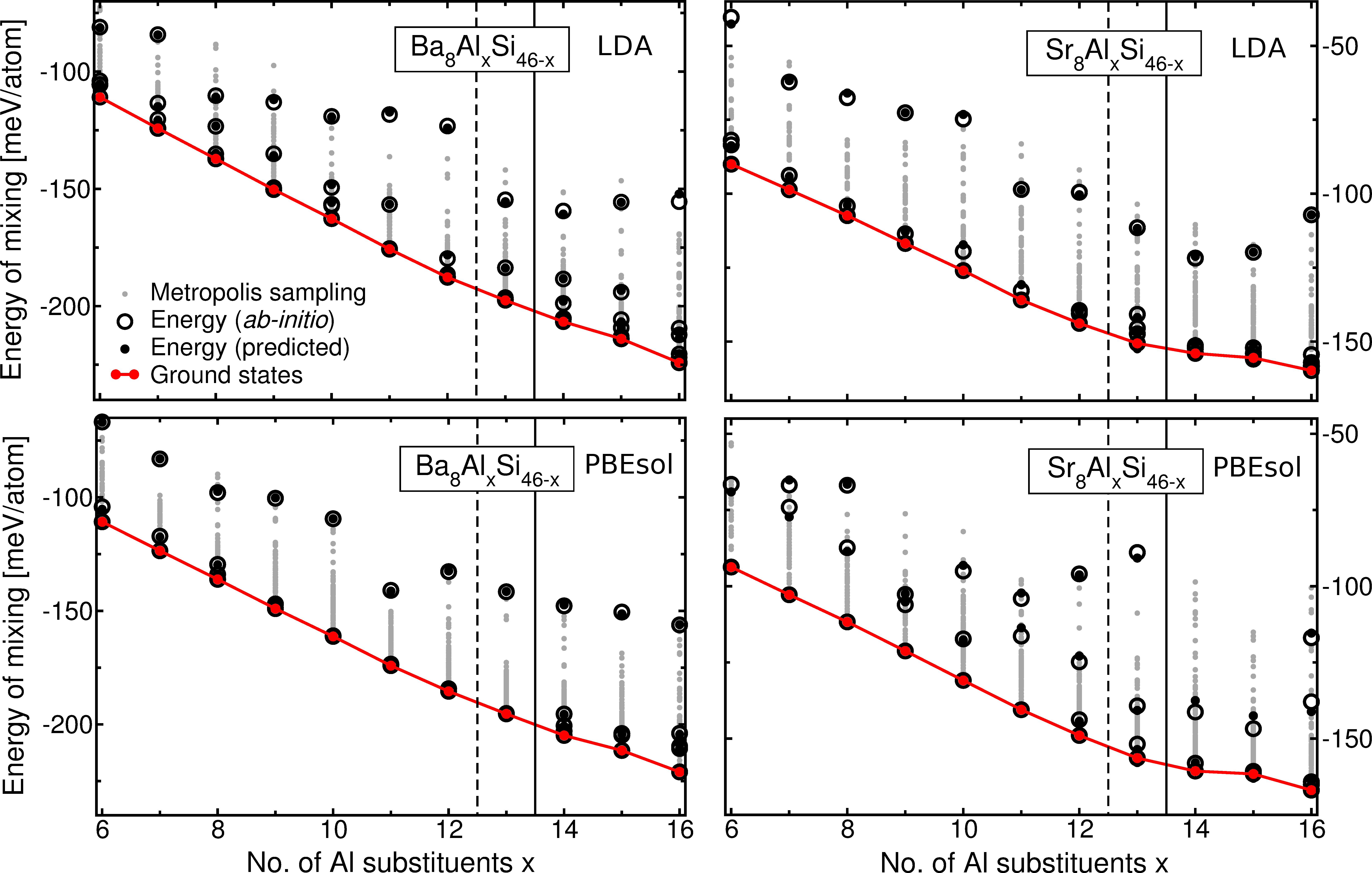}
  \caption{Ground-state search for \clath{Ba} (left) and \clath{Sr} (right) based on \ai 
  calculations with LDA (top) and PBEsol (bottom) as the exchange-correlation functional. 
  Two CEs are performed: CE1 in the range $x \in [6,13]$ and CE2 in the range $x \in [13,16]$.}
  \label{fig:GSSfinal}
 \end{figure}

The results of the iterative CE are shown in Fig.~\ref{fig:GSSfinal} for \clath{Ba} and \clath{Sr}. They are based on DFT calculations using LDA and PBEsol. The figure depicts the energy of mixing versus the number of Al substituents, $e(x)$ per atom, defined as
 \begin{equation}
 e(x) = \frac{1}{N}
 \left\lbrace 
 	E(x) - 
 	\left[ 
 		\widehat{E}(x=0) + 
 		\left( 
 			\widehat{E}(x=46) - \widehat{E}(x=0) 
 		\right) \frac{x}{46}
 	\right]
 \right\rbrace\,\textrm{.}\label{eq:emix}
 \end{equation}
This expression represents the energy difference between the \ai energy $E(x)$ and the energy obtained from a linear interpolation between the predicted energies of the pristine structures, \textit{e.g.} Ba$_{8}$Si$_{46}$ ($x$=0) and Ba$_{8}$Al$_{46}$ ($x$=46), taking the Ba clathrate as an example. $N$ denotes the number of atoms within the unit cell ($N$=54). The \ai results $e(x)$ are shown as black circles. Starting with the LDA results, for both clathrate compounds, the GS configurations decrease almost linearly with the Al content $x$ (red line), with an abrupt change in slope around  $x=13$. This behavior can be traced back to the electronic properties as will be discussed below. To accurately model this behavior, we perform two separate CEs by dividing the composition range into two sets of data. The first one, termed CE1, is based on the \ai data for $x \in [6,13]$ (to the left of the black solid line) and the second one (CE2) for $x \in [13,16]$ (to the right of the black dashed line). In both ranges, the predicted energies $\widehat{e}(x)$ (black dots) match well with their \ai energies (circles). [$\widehat{e}(x)$ can be obtained from Eq.~\ref{eq:emix} when replacing $E(x)$ by $\widehat{E}(x)$.] At $x$=13, predictions with CE1 or CE2 are equally good. In Fig.~\ref{fig:GSSfinal} predictions at this composition are obtained with CE2. With a well converged CE, a large Metropolis sampling (gray dots) is performed in order to search for new GS configurations and confirm those already found. In every Metropolis run, typically $5\times10^5$ configurations are sampled. For these samplings, a Boltzmann distribution with a temperature of $T=1000\,$K is used.

A fully independent ground-state search was performed with the PBEsol functional (bottom panels of Fig. \ref{fig:GSSfinal}). For both the Ba and Sr clathrates, the two functionals, LDA and PBEsol, lead to the same ground-state configurations indicating the robustness of the results.

 \subsection{Cluster expansion models}
 \label{sec:gss2}

To obtain the CE models in each iteration of the iCE, we employ an optimization procedure as described in Ref. \citenum{wa2002:atat}. A set of clusters with size  $N_c$ is defined by two parameters, the largest number of points $N_p^{max}$ and the maximum radius $R^{max}$, i.e. the maximum distance between two points in the cluster. All clusters with $N_p \le N_p^{max}$ and $R \le R^{max}$ are thus included in the set. A nested loop with increasing $R^{max}$ and $N_p^{max}$ is then performed, implying increasingly larger set sizes. The set leading to the smallest CV is selected. According to this optimization procedure, a small set of relevant clusters is identified. For this set, we finally perform a combinatorial optimization where all subsets of clusters are considered. As the number of relevant clusters is small, i.e. $N_c<N_t$, we set the penalization strength $A_{\alpha}$ in Eq.~\ref{eq:CVpenal} to zero.
  
\begin{figure}[h]
\includegraphics[scale=0.2]{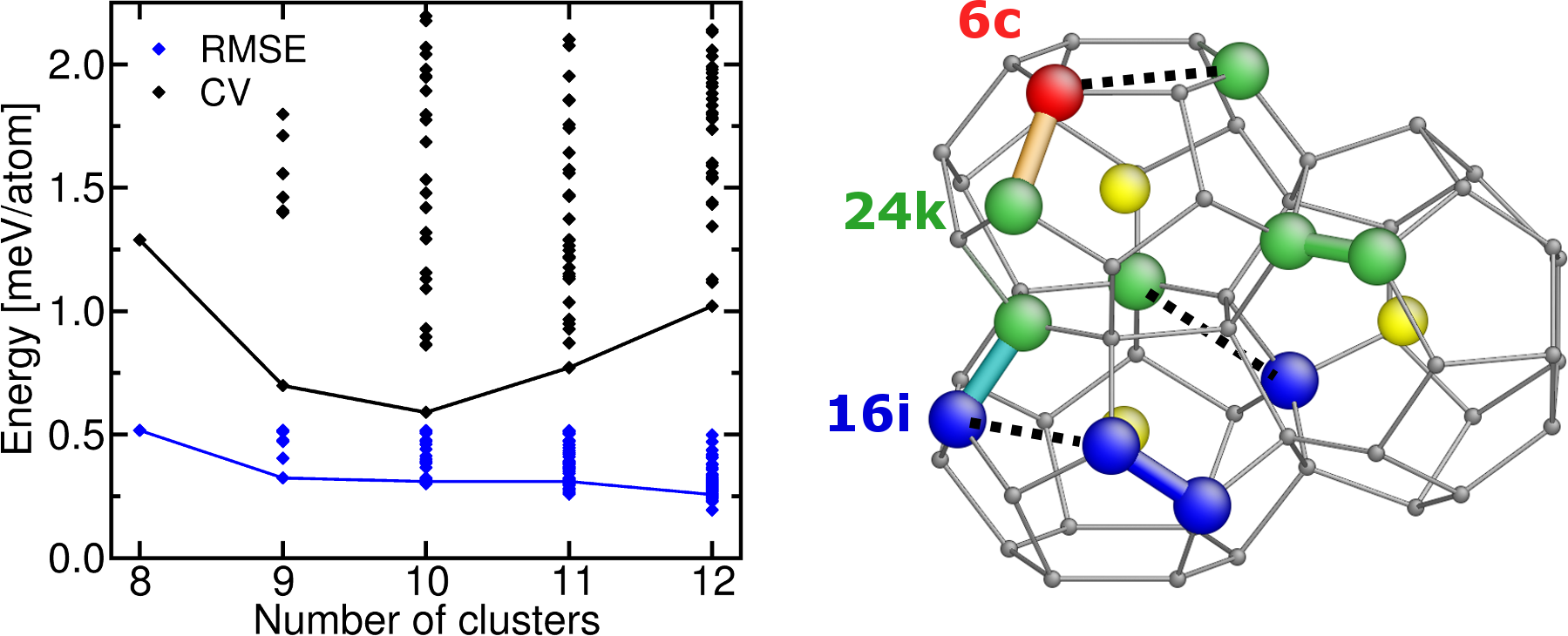}
\caption{(a) CV and RMSE versus the number of clusters, $N_c$, for CE2 (PBEsol) of \clath{Ba}. (b) Clusters leading to optimal CV. Guest atoms in yellow, host sites in grey; dashed lines indicate next-nearest neighbor two-point clusters.  Green color is used for $kk$ bonds, cyan for $ki$, orange for $kc$, and blue for $ii$.}
\label{fig:cluspict}
\end{figure}

\begin{table}[h]
\caption{Cluster expansion for \clath{Ba} corresponding to Fig.~\ref{fig:GSSfinal} (LDA and PBEsol). $N_t$ indicates the number of \ai values to perform the CE, $N_c$ the number of clusters. The ECIs are given in meV, except $J_0$ and $J_c$, that are given in eV. CV and RMSE are in meV/atom. In the indices, $k$, $i$, and $c$ stand for the Wyckoff sites $24k$, $16i$, and $6c$, respectively.}
\begin{tabular}{|| c || c | c || c | c ||}
 \hline \hline
  & \multicolumn{4}{c||}{\textbf{\clath{Ba}}}  \\ \hline
 & \multicolumn{2}{c||}{\textbf{LDA}} &
\multicolumn{2}{c||}{\textbf{PBEsol}} \\ \hline
 & CE1 & CE2 & CE1 & CE2 \\ \hline
 $N_t$ & 35 & 23 & 25 & 17  \\
 $N_c$ & 8 & 10 & 8 & 10  \\  \hline
 $J_0$ & -2132317.4541 & -2132320.4260 & -2133568.7971 & -2133571.8115 \\
\hline
\phantom{m} $J_k-J_c$ \phantom{m} & 239.4 & 217.8 & 262.3 & 206.4 \\
 $J_i-J_c$ & 316.0 & 325.1 & 342.8 & 284.3  \\
 $J_c$ & 1281.7591 & 1281.9953 & 1282.9376 & 1283.2045  \\ \hline
 $J_{kk}$ & 182.4 & 287.9 & 266.0 & 231.6  \\
 $J_{ii}$ & 456.0 & 393.5 & 348.4 & 380.5  \\
 $J_{ki}$ & 248.9 & 273.8 & 244.6 & 294.4  \\
 $J_{kc}$ & 337.1 & 287.8 & 282.3 & 276.3 \\ \hline
 $J_{k-i}$ & - & -12.5 & -  & - \\
 $J_{i-i}$ & - &  -  &  - & 33.1  \\
 $J_{k-c}$ & - & 20.5 & - & 11.3   \\ \hline
 CV & 1.43/{\bf 1.35} & 1.45/{\bf 1.08} & 1.13/{\bf 1.20} & 0.59/{\bf
0.60} \\
 RMSE & 1.04 & 0.66 & 0.69 & 0.31 \\
\hline \hline
 \end{tabular}
 \label{tab:CEall_Ba}
 \end{table}  
In  Fig.~\ref{fig:cluspict}(a) we demonstrate this combinatorial optimization for \clath{Ba} in the range $x \in [13,16]$ (CE2) for PBEsol. Each point represents a different cluster set. The CV (black diamonds) and the root mean-squared error (RMSE, blue diamonds) as a function of $N_c$ are depicted. The black solid line connects the sets yielding the lowest CV for each $N_c$, while the blue solid line connects the corresponding RMSE. The RMSE steadily decreases with the number of clusters, indicating an improvement of the fit to the \ai data. In contrast, the CV reaches a minimum at $N_c=10$, indicating the optimal CE (see Table \ref{tab:CEall_Ba}). The increase of the CV for $N_c>10$ reveals overfitting, i.e. fitting to noise in the \ai data. The optimal CE comprises the following 10 clusters: the empty cluster ($\alpha=0$); the three 1-point clusters arising from the three Wyckoff sites ($24k$, $16i$, and $6c$), labelled as $\alpha=k$, $i$, and $c$; and the four 2-point clusters $\alpha=kk$, $ki$, $kc$, and $ii$ (see Fig.~\ref{fig:cluspict}(b)). Additionally, two 2-point clusters consisting of next-nearest neighbor sites are present. These are denoted as $i$-$i$ and $k$-$c$ (dashed lines in Fig.~\ref{fig:cluspict}(b)). In the case of \clath{Sr}, the addition of next-nearest neighbor interactions lead to overfitting and thus are not included. 

Tables \ref{tab:CEall_Ba} and \ref{tab:CEall_Sr} provide summaries of the CEs for the two clathrate compounds. Two values for the CV are given in each case. The one on the left gives the average prediction error for the whole data set, as defined in Eq.\ref{eq:CVloo}. The one on the right (bold) represents the respective value for all configurations with an energy range from the ground-state energy to $20$meV/atom above. This latter quantity represents the accuracy of the predictions for low-energy configurations better. 

\begin{table}[h]
\caption{Same as Table \ref{tab:CEall_Ba} but for \clath{Sr}.}
\begin{tabular}{|| c || c | c || c | c ||}
 \hline \hline
  & \multicolumn{4}{c||}{\textbf{\clath{Sr}}} \\ \hline
 & \multicolumn{2}{c||}{\textbf{LDA}} &
\multicolumn{2}{c||}{\textbf{PBEsol}} \\ \hline
 & CE1 & CE2 & CE1 & CE2 \\ \hline
 $N_t$ & 28 & 18 & 25 & 19 \\
 $N_c$ & 8 & 8 & 8 & 8 \\  \hline
 $J_0$ & -1053234.7474 & -1053237.1841 & -1054204.5697 & -1054207.0499 \\
\hline
 $J_k-J_c$ & 283.5 & 301.1 & 283.6 & 287.9 \\
 $J_i-J_c$ & 323.0 & 372.1 & 339.2 & 357.7 \\
 $J_c$ & 1281.9361 & 1282.1087 & 1283.1146 & 1283.3056 \\ \hline
 $J_{kk}$ & 265.7 & 150.5 & 278.1 & 121.5 \\
 $J_{ii}$ & 296.4 & 327.0 & 299.9 & 203.8 \\
 $J_{ki}$ & 213.6 & 163.2 & 200.2 & 223.1 \\
 $J_{kc}$ & 258.8 & 259.8 & 235.1 & 243.4 \\ \hline
 CV & 1.94/{\bf 1.39} & 5.15/{\bf 1.64} & 3.84/{\bf 1.41} & 8.35/{\bf
1.13} \\
 RMSE &  1.16 & 0.96 & 1.47 & 1.74 \\
\hline \hline
 \end{tabular}
  \label{tab:CEall_Sr}
 \end{table}

 \section{Structural properties}
 \label{sec:struc}

 \subsection{Ground-state configurations} 
 \label{sec:struc1}

 \begin{table}[h]                         
 \caption{Site occupancies, represented by the tuple $(n_k,n_i,n_c)$, and number of bonds between Al atoms, $N_b$, for the GS configurations of different Al content $x$ in \clath{Ba} and \clath{Sr}, obtained by LDA and PBEsol, respectively.
}
\begin{tabular}{ | c || c | c || c | c |}
 \hline 
  & \multicolumn{2}{c||}{\textbf{\clath{Ba}}} & \multicolumn{2}{c|}{\textbf{\clath{Sr}}} \\ \hline
 $\mathbf{x}$ & \textbf{LDA} & \textbf{PBEsol} & \textbf{LDA} & \textbf{PBEsol} \\ \hline
 6 & (0,0,6)\,0 & (0,0,6)\,0 & (0,0,6)\,0 & (0,0,6)\,0 \\ \hline
 7 & (0,1,6)\,0 & (0,1,6)\,0 & (0,1,6)\,0 & (0,1,6)\,0 \\ \hline
 8 & (0,2,6)\,0 & (0,2,6)\,0 & (0,2,6)\,0 & (0,2,6)\,0 \\ \hline
 9 & (0,3,6)\,0 & (0,3,6)\,0 & (0,3,6)\,0 & (0,3,6)\,0 \\ 
   & (4,0,5)\,0 & (4,0,5)\,0 &            &            \\ \hline
 10 & (0,4,6)\,0 & (0,4,6)\,0 & (0,4,6)\,0 & (0,4,6)\,0 \\
    & (4,1,5)\,0 & (4,1,5)\,0 &         &         \\ \hline
 11 & (0,5,6)\,0 & (0,5,6)\,0 & (0,5,6)\,0 & (0,5,6)\,0 \\ 
    & (4,2,5)\,0 & (4,2,5)\,0 &         &         \\ \hline
 12 & (0,6,6)\,0 & (0,6,6)\,0 & (0,6,6)\,0 & (0,6,6)\,0 \\ 
    & (4,3,5)\,0 & (4,3,5)\,0 &         &         \\ \hline 
 13 & (4,4,5)\,0 & (4,4,5)\,0 & (0,7,6)\,0 & (0,7,6)\,0 \\
    & (0,7,6)\,0 & (0,7,6)\,0 &         &         \\
    & (8,1,4)\,0 & (8,1,4)\,0 &         &         \\ \hline
 14 & (8,2,4)\,0 & (8,2,4)\,0 & (0,8,6)\,0 & (0,8,6)\,0 \\ 
    & (0,8,6)\,0 & (0,8,6)\,0 &  &         \\ \hline
 15 &(12,0,3)\,0 &(12,0,3)\,0 & \,\,\,(1,8,6)\,1 & (1,8,6)\,1 \\
    &          &          & \,\,\,(2,8,5)\,0 & (2,8,5)\,0 \\ 
    &          &          & (12,0,3)\,0 & (12,0,3)\,0 \\ \hline
  16 &(12,1,3)\,0 &(12,1,3)\,0 & (12,1,3)\,0 & (12,1,3)\,0 \\
    &	   &          & (11,2,3)\,0	 & (11,2,3)\,0 \\ \hline \hline
 \end{tabular}
 \label{tab:GSconfig}
 \end{table}

From the CE we can deduce the site occupancies and numbers of Al-Al bonds per unit cell, which are shown in Table \ref{tab:GSconfig} for the GS configurations. These are represented by the notation $(n_k,n_i,n_c)\,N_b$, where $n_k$, $n_i$, and $n_c$ indicate the number of Al atoms sitting at the $24k$, $16i$, and $6c$ sites, respectively, and $N_{b}$ the number of bonds between Al atoms in the structure. A given tuple can represent several structures. Nevertheless, this simple representation is meaningful because it captures the main interactions. In addition to the ground-state configurations, the energies of a few quasi-degenerate structures (i.e., with energy differences below $\sim 2\,$meV/atom with respect to the GS) are present for some compositions. Overall, the results for \clath{Ba} and \clath{Sr} are very similar. More striking, for each of them the two functionals lead to exactly the same configurations. 

For both compounds, bonds between Al atoms are avoided over the entire compositional range with only one exception. This reflects the fact that all nearest-neighbor two-point interactions are positive (see Tabs. \ref{tab:CEall_Ba} and \ref{tab:CEall_Sr}). Moreover, the order of the one-point ECIs, $J_c < J_k < J_i $, indicates favored occupation of the $6c$ site, followed by a preference of the $24k$ and $16i$ sites, respectively. In line with this, the $6c$ site is fully occupied in the range $x \in [6,12]$ for \clath{Ba} and $[6,15]$ for \clath{Sr}. With further addition of Al atoms, the occupation of the $16i$ site increases while the $24k$ site remains empty. This seems in contradiction with the ordering of the single-point ECIs. However, adding an Al atom at a $24k$ site would lead to a structure with Al-Al bonds, since every $24k$ site has one $6c$ site as neighbor (see Fig.~\ref{fig:struc}). Thus the ECIs of the two-point clusters prevent occupation of the site $24k$ and rather cause occupation of the site $16i$. For instance, for \clath{Ba} (CE1, LDA) the energy gain for occupying a $16i$ position instead of $24k$ is $J_{kc} + J_{k} - J_{i} \sim 260.5$\,meV ($\sim 4.8$\,meV/atom).
  
In the intermediate composition range, $x \in [9,14]$, we observe a puzzling competition between single-point and two-point interactions for the Ba clathrate, leading to quasi-degenerate states through an interesting reordering. The occupation of the $6c$ site is reduced by one, increasing the occupancy of the $24k$ site, a neighbor of the now empty $6c$ position. The three remaining neighboring $24k$ sites are occupied by Al atoms at the expense of the $16i$ site. This {\it multiple exchange} can happen up to three times. The respective energy gain can be estimated from the ECIs as $4 J_k - 3 J_i - J_c$. For instance, for CE1 (LDA), this gives $9.6\,$meV ($\sim 0.2\,$meV/atom). For $x<13$, the GSs are represented by configurations $(0,x-6,6)\,0$. At $x=13$, the GS is $(4,4,5)\,0$ while two quasi-degenerate states, $(0,7,6)\,0$ and $(8,1,4)\,0$, are present. For $x=15$, the configuration $(12,0,3)\,0$ becomes the GS, which means half occupation of the $24k$ and $6c$ sites.

For the Sr clathrate, GS structures with a fully occupied $6c$ site are present up to $x=14$.  Avoidance of Al-Al bonds with full occupation of the $6c$ site is only possible up to this Al content. At this composition, the $16i$ site is half occupied, thus additional Al atoms at a $16i$ position would lead to an Al-Al bond between two $16i$ sites. At $x=15$, three degenerate structures appear. In contrast to all other cases, the GS contains one Al-Al bond. There exist, however, two more quasi-degenerate structures with $N_b$=0. For the charge-balanced composition ($x=16$), two quasi-degenerate configurations are present, with the GS being the same as for \clath{Ba}.

Prevention of bonds between trivalent elements (trivalent bonds, in short) has been reported from DFT calculations for other clathrate compounds with Ga and Ge atoms in the framework \cite{bl2001:str}. For the charge-balanced composition, the configuration $(12,1,3)\,0$ has been found as the most stable one out of nine different configurations.  From six samples of the clathrate K$_8$Al$_8$Si$_{38}$ the configuration $(2,0,6)\,0$ has been found lowest in energy\cite{he2014}. These results are in line with our findings for the Ba- and Sr-filled Al-Si based clathrates. It should be noted, though, that our method allows for exploring a much larger configurational and compositional space in an unbiased manner. In Ref. \citenum{chri2010:TE1} guidelines for the preferred occupancies have been derived postulating the avoidance of trivalent bonds. These rules are indeed confirmed by our results. An exception concerns the statement in Ref. \citenum{chri2010:TE1} that structures with $x=16$ without trivalent bonds do not exist. Conversely, we clearly show that the GS at the charge-balanced composition does not contain Al-Al bonds.

 \begin{figure}[h]
  \centering
  \includegraphics[scale=0.55]{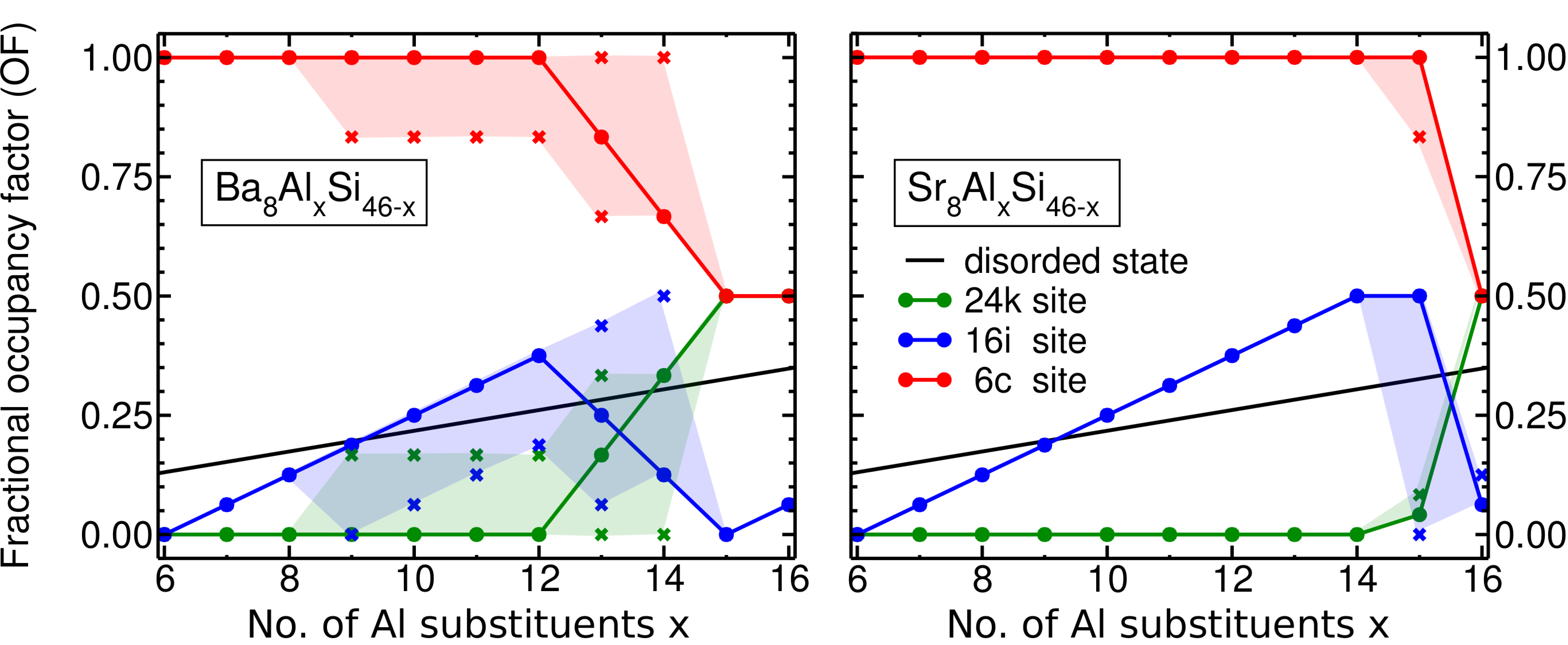}
  \caption{Fractional occupancy factors of the ground-state configurations (dots) and their degenerate states (crosses). The shaded areas indicate overall possible values.}
 \label{fig:occ}
 \end{figure}

Figure \ref{fig:occ} shows the fractional occupancy factors OF$=n_w/N_w$ of the ground-state configurations (dots) and their degenerate states (crosses), with $N_w$ being the multiplicity of Wyckoff site $w$. In a material, both the GS and the (nearly) degenerate states can be present simultaneously in different regions, therefore the average OF can take any value in between. This is indicated by the shaded areas. 

\subsection{Lattice constants}
\label{sec:struc2}
\begin{figure}[h]
  \centering
  \includegraphics[scale=0.55]{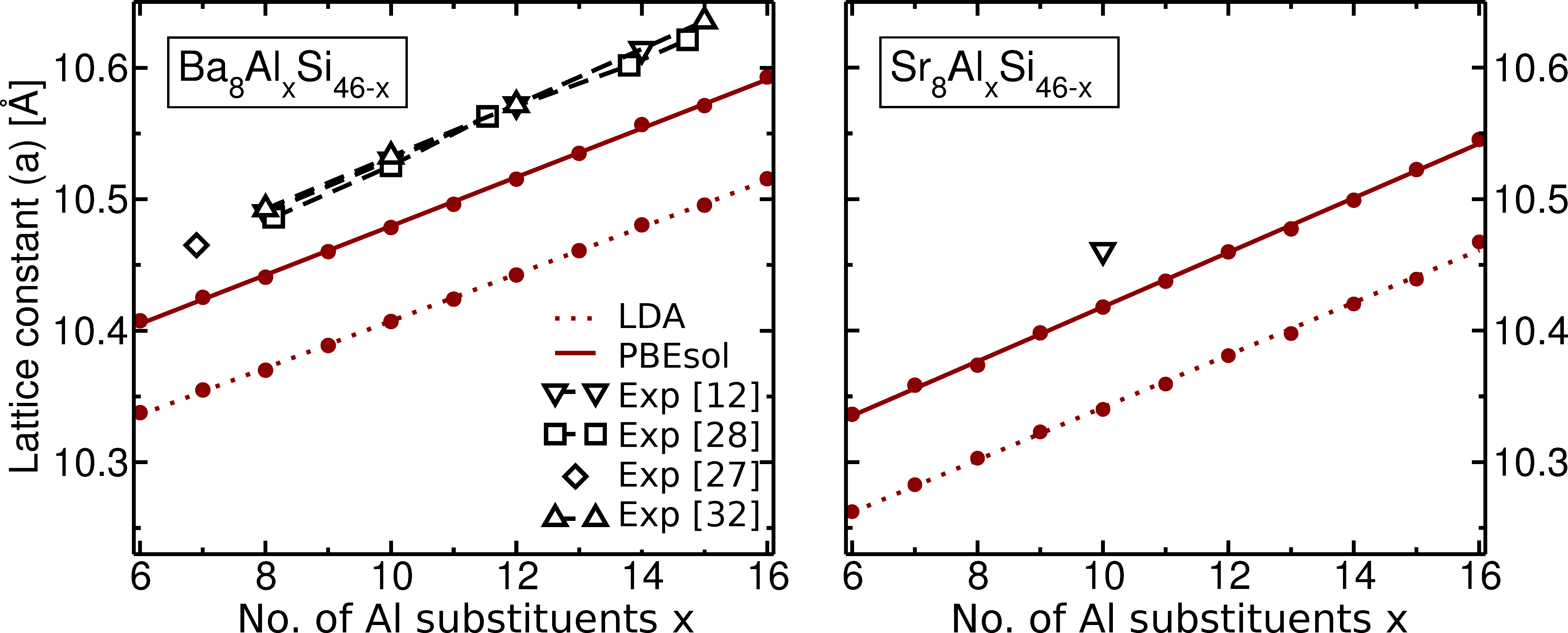}
  \caption{Lattice constant vs. the number of Al substituents for the GS
structures  of \clath{Ba} (left) and \clath{Sr} (right), obtained by LDA
(dotted lines) and PBEsol (solid lines). Experiments from Refs.
\citenum{ro2012:Srclath}, \citenum{ro2012:Baclath},
\citenum{bo2015:Bax7}, and \citenum{bomi2015:pc}.}
  \label{fig:latcon}
 \end{figure}

The lattice constants of the GS structures are presented in Fig.~\ref{fig:latcon} for \clath{Ba} (left) and \clath{Sr} (right). Both compounds reveal a linear increase with the number of Al substituents, as obtained with LDA (dotted lines) and PBEsol (solid lines). This is explained by the larger atomic radius of Al atoms as compared to Si \cite{wa1965:orbit}. The slope of $\Delta a / \Delta x \approx 0.018\,$\r{A} for \clath{Ba} is close to the experimental value of about 0.020\,\r{A} (black dashed lines) \cite{ro2012:Srclath,ro2012:Baclath}. \clath{Sr} exhibits a similar slope of $\Delta a / \Delta x \approx 0.020$ \r{A}. The smaller size of Sr atoms in comparison to Ba \cite{wa1965:orbit} leads to a smaller lattice spacing by about 0.07\,\r{A} in the entire composition range. This difference agrees well with the difference in the experimental values at $x=10$ (the only composition synthesized so far for \clath{Sr} \cite{ro2012:Srclath}). 

As expected, the lattice parameters obtained by LDA are smaller ($\sim 0.065\,$\r{A}) than those from PBEsol. The latter still underestimates the experimental value by about 0.5$\%$. While for solid Al and Si the relative error is less than 0.5\% for both functionals \cite{ha2009:func}, for alkaline-earth metals, as Sr and Ba, the lattice spacing is underestimated by LDA by more than 4\% and by PBEsol by more than 2\% \cite{ha2009:func}. This suggests for the clathrates that the slight systematic underestimation by PBEsol is caused by the guest atoms.

\subsection{Bond-distances}
\label{sec:struc3}
 
\begin{figure}[h!]
  \centering
  \includegraphics[scale=0.55]{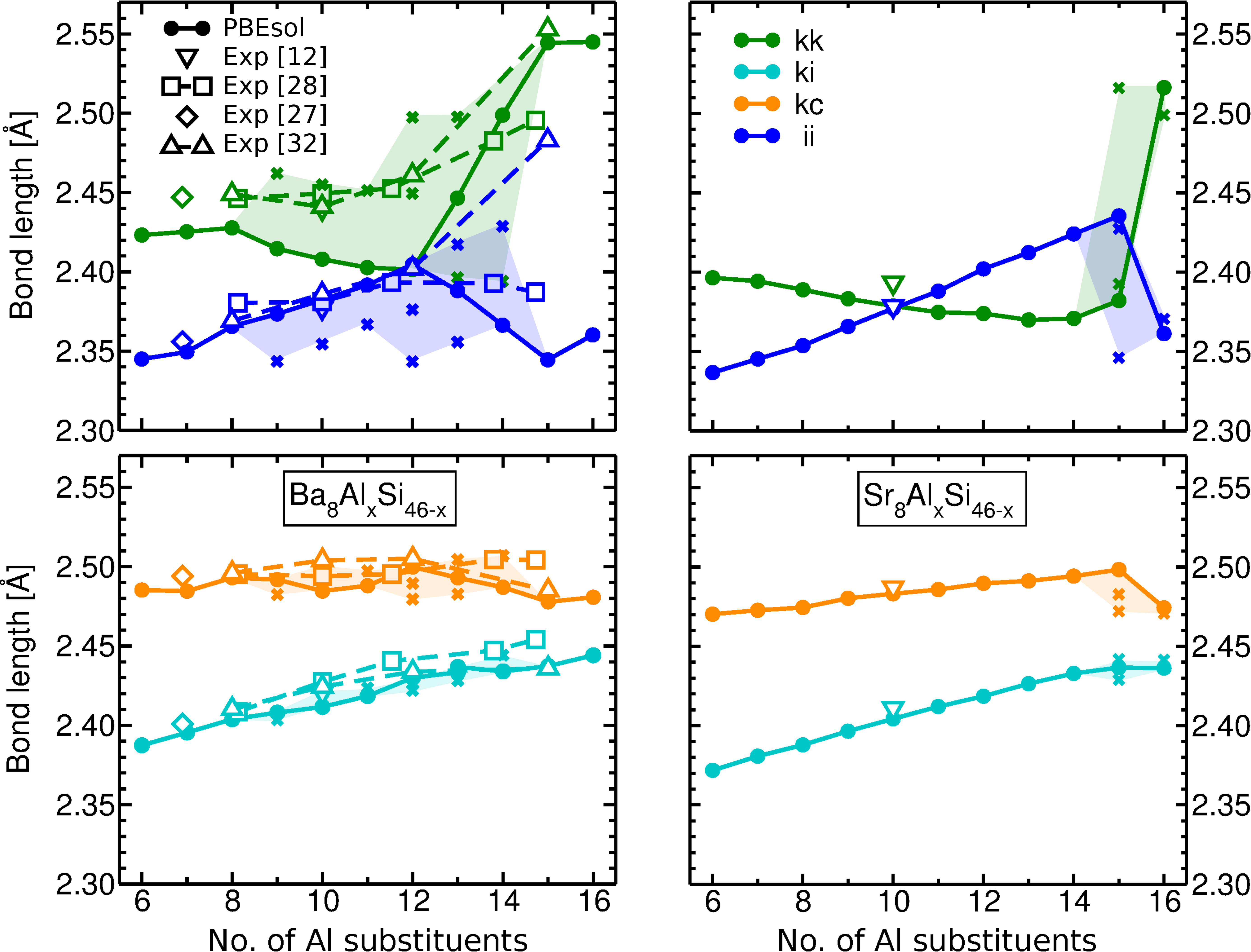}
  \caption{Average bond distances for the GS configurations of Ba (left)
and  Sr (right) clathrates (dots and solid lines are a guides to the eye)
and their degenerate states (crosses) as obtained by PBEsol. Shaded areas
indicate possible values of the bond distances, assuming a mixture of GS
and quasidegenerate configurations. Experimental data from Refs.
\citenum{ro2012:Srclath}, \citenum{ro2012:Baclath},
\citenum{bo2015:Bax7}, and \citenum{bomi2015:pc}.}
  \label{fig:bonddist}
\end{figure}

Like the lattice parameters, also the average bond distances reveal a dependence on the number of Al substituents, as can be seen in Fig. \ref{fig:bonddist}. The dots indicate the PBEsol data; the LDA results (not shown) are uniformly smaller by about 0.02\,\r{A}.

Focusing on the Ba clathrate, the $ii$ and $ki$ bonds show a monotonic, almost linear increase with Al content for $6 \leq x\leq 12$. This reflects the fact that in this composition range, additional Al atoms occupy the $16i$ sites, resulting in larger bond lengths when a $16i$ site is involved, i.e., the $ii$ and $ki$ distances. Since no $24k$ site is occupied in the range $x<12$, the $kk$ bond distance remains almost constant. As shown before (Table \ref{tab:GSconfig}), occupation of the $24k$ site appears from $x=13$ onwards leading to a drastic increase of the $kk$ bond and a corresponding decrease of the $ii$ bond from $x=12$ to $x=13$ due to simultaneous depletion of this site. This trend continues until the charge-balanced composition is reached, where the $24k$ site is half occupied in the GS configuration. The $kc$ bond length is almost independent of the Al content. The reason becomes clear by recalling the Al occupations of the $6c$ and $24k$ sites. At low $x$, full $6c$ and empty $24k$ sites imply that one Al atom is part of every $kc$ bond. At high $x$, the $6c$ position is less populated, but at the same time the surrounding $24k$ sites of an empty $6c$ site are occupied, thus, the number of Al atoms forming part of $kc$ bonds does not change.

For the Sr compound, the trends in the low substitutional range are analogous to those of the Ba counterpart. However, the linear increase of the $ii$ and $ki$ bond distances continues until the composition of $x=15$ is reached. This is again related to the preferred occupation of the $16i$ site. At $x=16$, the energetic ordering of the configurations changes abruptly (see above), and the bond distances now appear similar to those of the Ba clathrates. 

In Fig.~\ref{fig:bonddist} also experimental data are depicted. For the Ba case, the $kc$ and $ki$ values are in good agreement with our calculations for the entire composition range (including nearly degenerate states shown as crosses). The experimental $kk$ bond length is slightly larger than the theoretical value for $x=6$-$7$; this may point to a marginal presence of Al at the $24k$ site in the real sample. For $x \geq 9$, good agreement between theory and experiment is observed. In this composition range containing quasi-degenerate states, the experimental values lie inside the green shaded area. This area indicates the possible values of bond lengths assuming a mixture of GS and quasi-degenerate configurations. Concerning the $ii$ bond, almost all experimental values lie within the range of the calculated ones (blue shaded area). For $kk$ and $ii$, the bond lengths of the degenerate states are very different from those of the GS. Notably, also the experimental values largely deviate from each other. Regarding the Sr clathrate, there is good agreement with experiment. The slightly larger measured size of the $kk$ bond for $x=10$ may be of the same origin as in the Ba case.

\section{Electronic properties}
\label{sec:elect}

 \begin{figure}[h]
 \centering
 \includegraphics[scale=0.6]{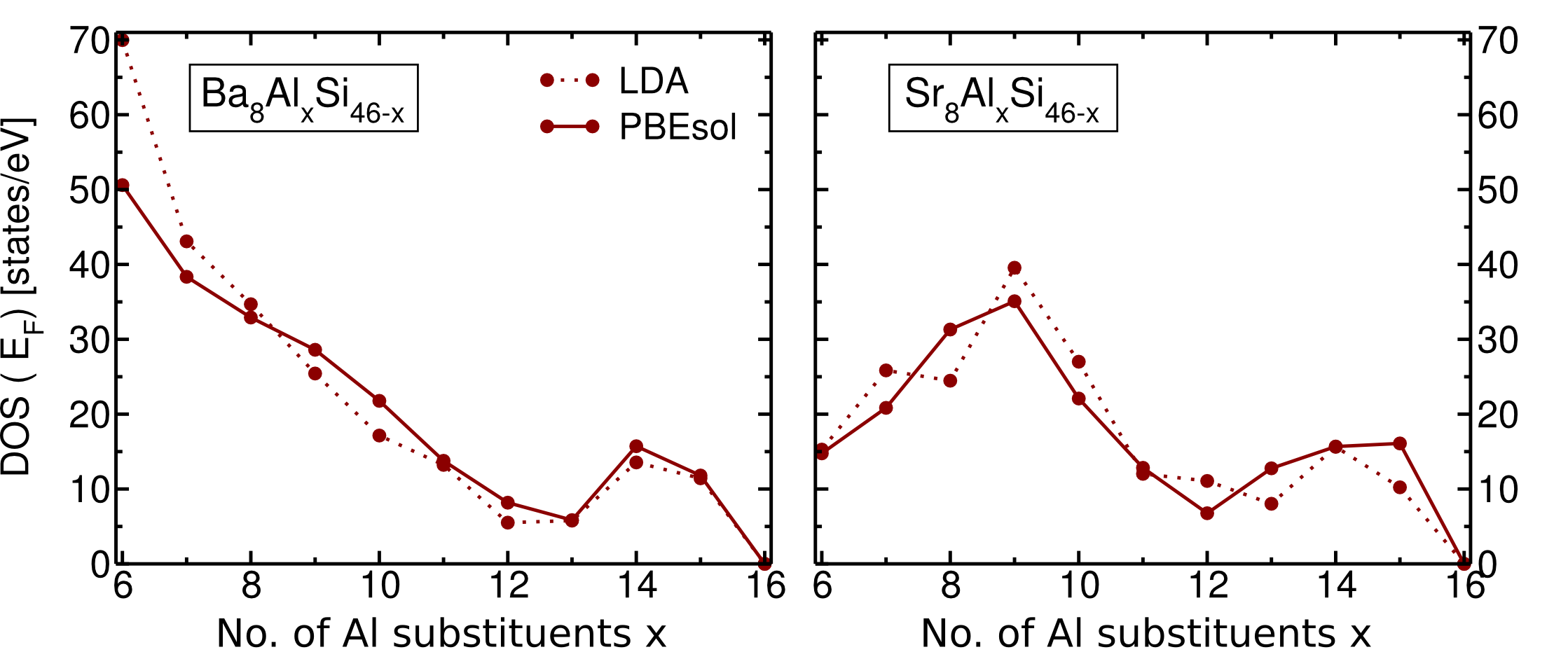}
 \caption{Density of states at the Fermi level, DOS(E$_F$), for the GS configurations of \clath{Ba} (left) and \clath{Sr} (right) as obtained from LDA calculations (dashed lines) and PBEsol (solid lines).}
 \label{fig:dosf}
 \end{figure}

The density of states at the Fermi level, DOS(E$_F$), is shown in Fig. \ref{fig:dosf} for the GS configurations of \clath{Ba} and \clath{Sr}. Metallic behavior is observed over the whole composition range, except at $x=16$, where it approaches zero. Thus, the charge-balanced composition, indeed, exhibits a semiconducting GS\footnote{Due to the electronic-entropy contribution, semiconducting behavior for the lowest-energy charge-balanced composition, may represent a disadvantage in terms of stability at finite temperatures, relative to metallic compositions. This could explain the difficulties in synthesizing the charge-balanced composition for Ba-filled clathrates, which could be achieved only up to $x=15$.}. In all cases, a local minimum close to $x=13$ points to a change in the electronic configuration of conduction electrons around this concentration. This goes hand in hand with the change in slope in the GS energy of mixing in Fig.~\ref{fig:GSSfinal}. In line with this observation is the finding that the ECIs obtained by splitting the \ai data set at $x=13$ to perform two cluster expansions, resulted in much better predictions than without doing so.

 \begin{figure}[h]
  \centering
  \includegraphics[scale=0.75]{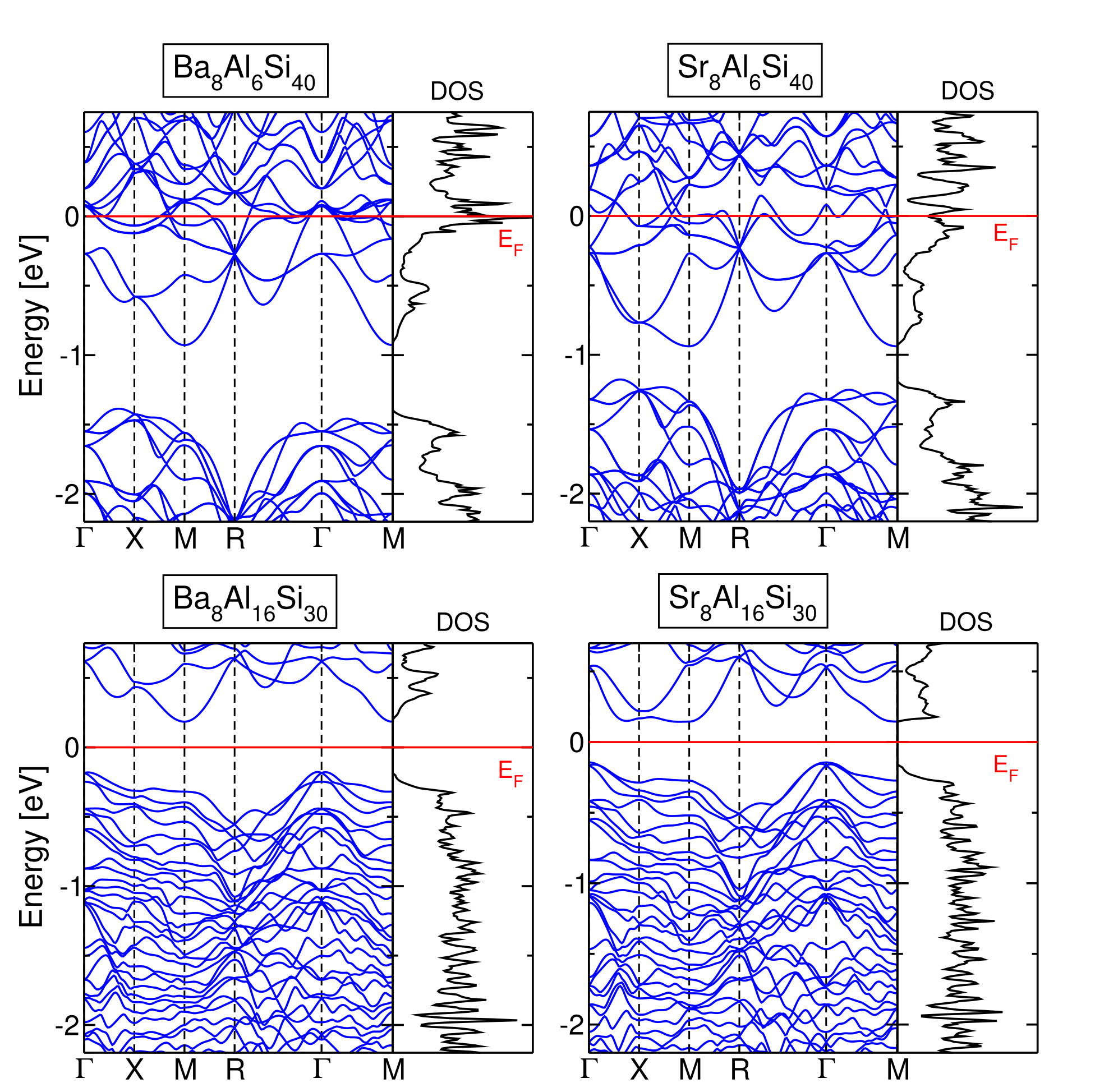}
  \caption{Band structure and density of states for Ba$_8$Al$_6$Si$_{40}$ (top left), Sr$_8$Al$_6$Si$_{40}$ (top right), 
  Ba$_8$Al$_{16}$Si$_{30}$ (bottom left), and Sr$_8$Al$_{16}$Si$_{30}$ (bottom right), as obtained with the PBEsol functional.}
  \label{fig:bandstr}
 \end{figure}

Band structures and densities of states of \clath{Ba} and \clath{Sr} for $x=6$ and $16$ are illustrated in Fig.~\ref{fig:bandstr}. 
There is a gap well below the Fermi level amounting to 0.25 eV for \clath{Sr} and 0.45 eV for \clath{Ba}, respectively, as obtained by PBEsol and in agreement with Ref. \citenum{ro2012:Srclath}. By increasing the Al concentration, corresponding to a removal of ``free'' electrons, this gap moves up in energy. At the charge-balanced composition, it is centered at the Fermi level and presents an indirect Kohn-Sham band gap along the $\Gamma$-M direction of 0.30 and 0.36 eV for \clath{Sr} and \clath{Ba}, respectively (see Table \ref{tab:dosZintl}).

\begin{figure}[h!]
  \centering
  \includegraphics[scale=0.55]{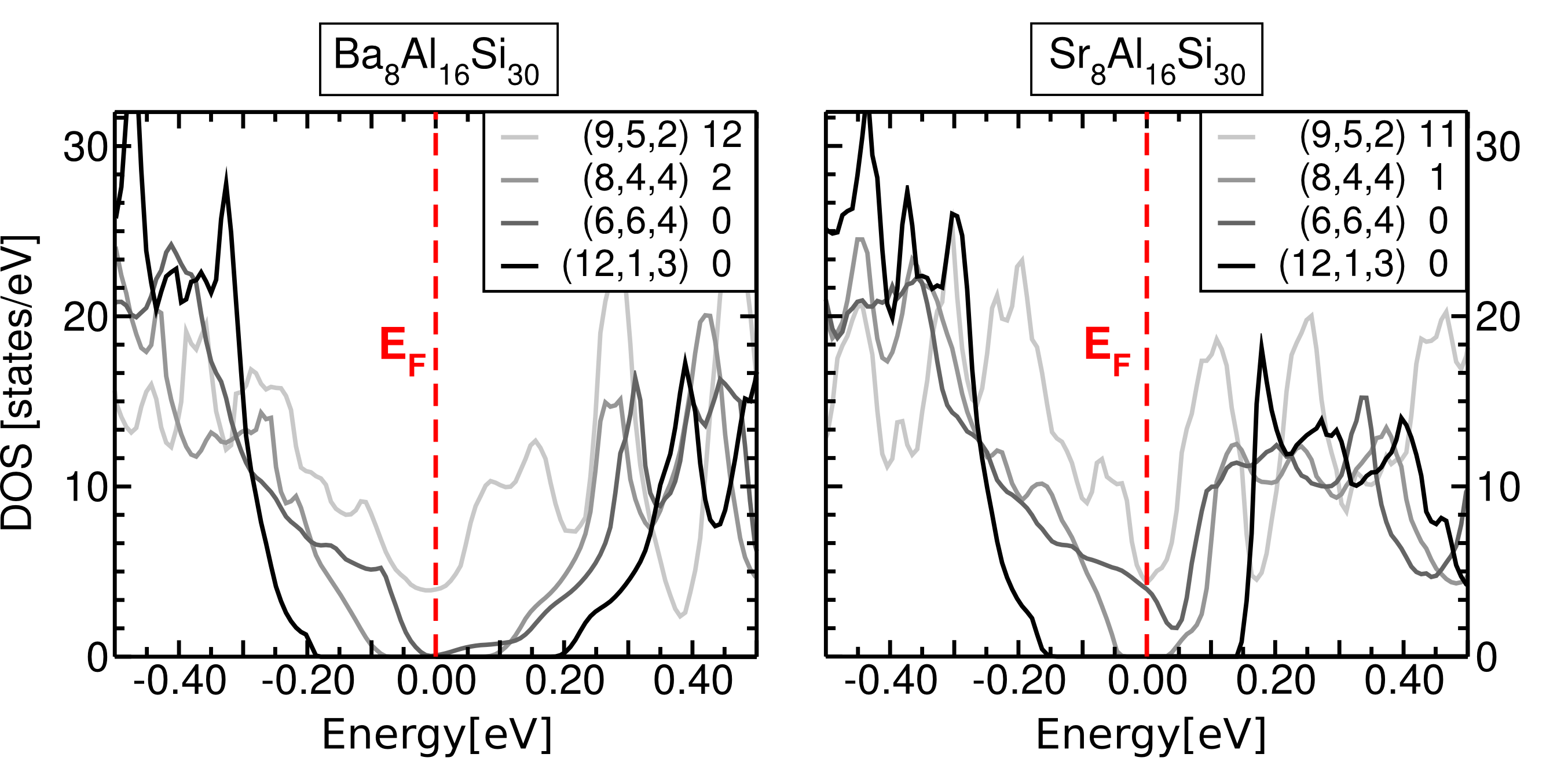}
  \caption{Density of states around E$_F$ for structures at the charge-balanced composition obtained with PBEsol for the Ba (left) and the Sr clathrate (right). For the nomenclature of the configurations, see Table \ref{tab:GSconfig}.}
  \label{fig:dosZ}
 \end{figure}

Since semiconducting behavior is beneficial for the thermoelectric properties, we have closer look at the electronic behavior of these materials at the charge-balanced composition. Figure \ref{fig:dosZ} illustrates the DOS around the Fermi level for Ba$_8$Al$_{16}$Si$_{30}$ (left) and Sr$_8$Al$_{16}$Si$_{30}$ (right) for different configurations appearing during the ground-state search with the PBEsol functional. Their energies of mixing and band gaps (if present) are summarized in Table \ref{tab:dosZintl}, showing also the corresponding LDA results. It is interesting to note that the configurations with a large number of Al-Al bonds clearly reveal metallic behavior. There exist also structures with a few Al-Al bonds and a small Kohn-Sham band gap [tuple (8,4,4)\,2]. The structures with no Al-Al bonds are, as already discussed, lower in energy, but not all of them show semiconducting behavior. However, for both materials the GS configuration has the largest Kohn-Sham band gap with respect to other configurations. 

The observation of a band gap in Sr$_8$Al$_{16}$Si$_{30}$ is in seeming contradiction with calculations reported in Ref. \citenum{ro2012:Srclath} giving metallic behavior for the same composition. However, this difference is explained by the facts that the  configuration considered in this reference is not the GS, and all 16 Al atoms were placed at $16i$ sites. This structure has $8$ Al-Al bonds, since each $16i$ site is linked to another $16i$ site. As noted above, Al-Al bonds lead to high-energy structures with vanishing or very small gaps. Semiconducting behavior of Ba$_8$Al$_{16}$Si$_{30}$ has already been reported for the non-GS configuration (9,4,3)\,0 \cite{nen2008:proBaclath}. From our CE we confirm this configuration to be close in energy to the GS.

 \begin{table}[h!]
\caption{Configurations at the charge-balanced composition, exhibiting either metallic or semiconducting behavior, together with their energy of mixing and Kohn-Sham band gap (if applicable). For the nomenclature of the configurations, see Table \ref{tab:GSconfig}.}
\begin{tabular}{ || c | c | c || c | c | c ||}
  \hline \hline
  \multicolumn{6}{||c||}{\textbf{\clath{Ba}}} \\ \hline 
  \multicolumn{3}{||c||}{\textbf{LDA}} & \multicolumn{3}{c||}{\textbf{PBEsol}} \\ \hline
  Configuration & $e_{\textrm{mix}}$ & $E_{g}$ & Configuration & $e_{\textrm{mix}}$ & $E_{g}$ \\ \hline
  \,\,\,\,(9,5,2)\,12 & -155.5 & - & \phantom{m}\,\,\,\,(9,5,2)\,12\phantom{m}& -156.2 & - \\
  \,\,(2,8,6)\,2  & -209.4 & - & \,\,(8,4,4)\,2 & -210.6 & 0.16  \\
  \,\,(8,4,4)\,2  & -216.8 & - & \,\,(6,6,4)\,0 & -215.5 & -   \\
  (12,1,3)\,0 & -224.1 & 0.34 & (12,1,3)\,0 & -220.9 & 0.36  \\ \hline \hline
  \multicolumn{6}{||c||}{\textbf{\clath{Sr}}} \\ \hline
  \multicolumn{3}{||c||}{\textbf{LDA}} & \multicolumn{3}{c||}{\textbf{PBEsol}} \\ \hline
  Configuration & $e_{\textrm{mix}}$ & $E_{g}$ & Configuration & $e_{\textrm{mix}}$ & $E_{g}$ \\ \hline
 \,\,\,\,(9,5,2)\,11 &  -107.1 & - & \,\,\,\,(9,5,2)\,11 & -116.9 & -  \\
 \,\,(3,8,5)\,1  &  -154.4 & - & \,\,(8,4,4)\,1  & -164.2 & 0.08  \\
 \,\,(6,6,4)\,0  &  -157.3 & - & \,\,(6,6,4)\,0  & -165.0 & -  \\
 (12,1,3)\,0 &  -159.8 & 0.30 & (12,1,3)\,0 & -166.9 & 0.30 \\
 \hline \hline
 \end{tabular}
 \label{tab:dosZintl}
 \end{table}

In general, the electronic structure of these materials strongly depends on the atomic configuration. The predicted semiconducting behavior of both compounds in some configurations at the charge-balanced composition, and predominantly those of the ground states, is an exciting finding in view of their application in thermoelectric devices.

\section{Conclusions and outlook}
 \label{sec:conc}
In the present work, we have investigated the clathrate compounds \clath{Ba} and \clath{Sr} ($6 \leq x \leq 16 $) with respect to their structural stability and electronic properties. Their large configurational space has been explored with \spel{CELL} \cite{cell}, an iterative cluster expansion technique for large parent cells. Thanks to its numerical efficiency we have been able to predict ground-state configurations as a function of Al content without making any presuppositions. For both compounds, these GS have been verified by two independent sets of \ai calculations based on the exchange-correlation functionals LDA and PBEsol, respectively. The CE yields total energies with an accuracy of less than $2\,$meV/atom for low lying configurations. This accuracy has allowed us to determine site preferences and bond lengths. Detailed analysis has revealed that Al-Al bonds are energetically unfavorable. There is a clear preference for Al atoms to occupy the $6c$ site. In the Ba compound, a transition from a full $6c$ to half $6c$ occupation takes place between $x$=10 and 14, and the $6c$ site remains half occupied up to the charge-balanced composition. The GS configurations of the Sr clathrate reveal a constant increase in the occupation of the $16i$ site up to $x=15$ and a drastic configurational rearrangement at $x=16$. Both clathrate types exhibit the same GS configuration at the charge-balanced composition, with 12 atoms at the $24k$ site (half occupancy), one atom at the $16i$ site, and 3 atoms at the $6c$ site (half occupancy),  thereby completely avoiding Al-Al bonds. The identification of such a configuration as the GS is an important finding, as it reveals semiconducting behavior. Other higher-energy configurations at charge-balanced composition are metallic, indicating that  structural ordering results in a metal-semiconductor transition. 

We have confirmed the linear increase of the lattice parameter with the number of Al substituents observed for \clath{Ba}, as well as the changes in the bond distances revealed by various experiments. The bond distances have been found to be strongly correlated to the Al occupation of the Wyckoff sites as reported by previous studies \cite{ro2012:Baclath,bo2015:Bax7}. Since in X-ray diffraction experiments Al and Si atoms are not clearly distinguishable \cite{chri2010:TE1,ro2012:Srclath,ro2012:Baclath,bo2015:Bax7}, a reliable model for describing the correlations between the site occupancies and the bond distances is highly desired. Our findings may form a solid basis for building such a model. An extension of our work towards finite-temperature effects will be subject of a forthcoming paper. Considering the large influence of the configurational ordering on the investigated quantities, this step is promising for a quantitative analysis of experimental observations.

Input and output files of our calculations can be downloaded from the NOMAD repository by following the link in Ref. \citenum{mt2017data}.

\section{Acknowledgements}
Work supported by the Einstein Foundation Berlin (project ETERNAL). The Norddeutsche Verbund f\"ur Hoch- und H\"ochstleistungsrechnenn (HLRN) enabled the computationally involved \ai calculations. We appreciate discussions with Bodo B\"ohme, Michael Baitinger, and Yuri Grin (MPI, Dresden), and thank them for providing us with experimental data prior to publication. Partial funding from the European Union's Horizon 2020 research and innovation programme, grant agreement No. 676580 through the Center of Excellene NOMAD (Novel Materials Discovery Laboratory, https://NOMAD-CoE.eu) is acknowledged. 


\providecommand{\latin}[1]{#1}
\makeatletter
\providecommand{\doi}
  {\begingroup\let\do\@makeother\dospecials
  \catcode`\{=1 \catcode`\}=2 \doi@aux}
\providecommand{\doi@aux}[1]{\endgroup\texttt{#1}}
\makeatother
\providecommand*\mcitethebibliography{\thebibliography}
\csname @ifundefined\endcsname{endmcitethebibliography}
  {\let\endmcitethebibliography\endthebibliography}{}

\end{document}